\documentclass[10pt,a4paper,onecolumn]{article} 

\usepackage[pdftex]{graphicx}
\usepackage{amssymb}
\usepackage{amsmath}
\usepackage[square,comma,numbers,sort&compress]{natbib}
\usepackage{url}
\usepackage[font={small}, labelfont=bf]{caption} 

\title{Highly comparative time-series analysis: The empirical structure of time series and their methods}

\author
{Ben D. Fulcher$^{1}$, Max A. Little$^{1}$, Nick S. Jones$^{1,2}$\\
\\
\normalsize{$^{1}$Department of Physics, University of Oxford}\\
\\
\normalsize{$^{2}$Department of Mathematics, Imperial College, London}\\
}

\date{}

\begin{document}

\maketitle

\begin{abstract}
The process of collecting and organizing sets of observations represents a common theme throughout the history of science.
However, despite the ubiquity of scientists measuring, recording, and analyzing the dynamics of different processes, an extensive organization of scientific time-series data and analysis methods has never been performed.
Addressing this, annotated collections of over 35\,000 real-world and model-generated time series and over 9\,000 time-series analysis algorithms are analyzed in this work.
We introduce reduced representations of both time series, in terms of their properties measured by diverse scientific methods, and of time-series analysis methods, in terms of their behaviour on empirical time series, and use them to organize these interdisciplinary resources.
This new approach to comparing across diverse scientific data and methods allows us to organize time-series datasets automatically according to their properties, retrieve alternatives to particular analysis methods developed in other scientific disciplines, and automate the selection of useful methods for time-series classification and regression tasks.
The broad scientific utility of these tools is demonstrated on datasets of electroencephalograms, self-affine time series, heart beat intervals, speech signals, and others, in each case contributing novel analysis techniques to the existing literature.
Highly comparative techniques that compare across an interdisciplinary literature can thus be used to guide more focused research in time-series analysis for applications across the scientific disciplines.
\end{abstract}

\section{Introduction} \label{sec:introduction}
Time series, measurements of a quantity taken over time, are fundamental data objects studied across the scientific disciplines, including measurements of stock prices in finance, ion fluxes in astrophysics, atmospheric air temperatures in meteorology, and human heart beats in medicine.
In order to understand the structure in these signals and the mechanisms that underly them, scientists have developed a large variety of techniques: methods based on fluctuation analysis are frequently used in Physics \cite{Peng95a}, generalized autoregressive conditional heteroskedasticity (GARCH) models are common in Economics \cite{Bollerslev86}, and entropy measures like Sample Entropy are popular in medical time-series analysis \cite{Richman00}, for example.
But are these methods that have been developed in different disciplinary contexts summarizing time series in unique and useful ways, or is there something to be learned by synthesizing and comparing them?
In this paper, we address this question by assembling extensive annotated libraries of both time-series data, and methods for time-series analysis.
We use each to organize the other: methods are characterized by their behaviour across a wide variety of different time series, and time series are characterized by the outputs of diverse scientific methods applied to them.
The structure that emerges from these collections of data and methods provides a new, unified platform for understanding the interdisciplinary time-series analysis literature.

Although a strong theoretical foundation underlies many time-series analysis methods, their great number and interdisciplinary diversity makes it very difficult to determine how methods developed in different disciplines relate to one another, and for scientists to select appropriate methods for their finite, noisy data.
It is similarly difficult to determine how time series studied in one scientific discipline compare to those studied in other disciplines, or to the dynamics defined by theoretical time-series models.
The automated, data-driven techniques introduced in this work address these practical difficulties by allowing diverse methods and data from across the sciences to be compared in a unified representation.
The high-throughput analysis of the DNA microarray in biology is now a standard complement to traditional, more focused research efforts; we envisage the comparative techniques introduced in this work providing a similarly guiding role for scientific time-series analysis.
Previous comparisons of time-series analysis methods have been performed only in specific disciplinary contexts and on a small scale, and attempts to organize large time-series datasets have typically involved time series of a fixed length and measured from a single system \cite{Liao05, Wang12}.
This work therefore represents the first scientific comparison of its kind and is unprecedented in both its scale and interdisciplinary breadth.

The paper is structured as follows.
In Sec. \ref{sec:framework}, we describe our extensive, annotated libraries of scientific time-series data and time-series analysis methods and introduce our computational framework.
Then, in Sec. \ref{sec:empirical_structure}, we show how these diverse collections of both methods (Sec. \ref{sec:empirical_structure_methods}) and data (Sec. \ref{sec:empirical_structure_timeseries}) can be organized meaningfully: by representing methods using their behaviour on the data, and representing time series by their measured properties.
Using this new ability to structure diverse collections of time series and their methods, we introduce a range of useful techniques, including the ability to connect specific pieces of time-series data to similar real-world and model-generated time series, and to link a specific time-series analysis methods to a range of alternatives from across the literature.
The highly comparative analysis techniques are demonstrated using scientific case studies in Sec.~\ref{sec:applications}.
The diverse range of scientific methods in our library are used to contribute new insights into existing time-series analysis problems: uncovering meaningful structure in time-series datasets, and selecting relevant methods for general time-series classification and regression tasks automatically.
We contrast our approach, that compares across a wide, interdisciplinary time-series analysis literature, to that of conventional studies that focus on small sets of manually-selected techniques with minimal comparison to alternatives.
Our main conclusions are summarized in Sec. \ref{sec:conclusions}.

\section{Framework}\label{sec:framework}
We assembled annotated libraries of: (i) 38\,190 univariate time series measured from diverse real-world systems and generated from a variety of synthetic model systems, and (ii) 9\,613 time-series analysis algorithms developed in a range of scientific literatures, as well as new analysis methods developed by us in this work.
Over 20\,000 real-world time series were acquired, primarily from publicly-available databases, which have been selected to encompass a representative sample of the types of signals measured in scientific disciplines: from meteorology (e.g., temperature, air pressure, rainfall, river flow), medicine (e.g., heart-beat intervals, electrocardiograms, electroencephalograms), audio (e.g., human speech, music, sound effects, animal sounds), astrophysics (e.g., solar radio flux, interplanetary magnetic field), finance (e.g., exchange rates, stock prices), and others.
In addition, we generated over 10\,000 synthetic time series, including outputs from a range of nonlinear and chaotic maps, various dynamical systems/flows, correlated noise models, and other stochastic processes.
Although it is clearly unfeasible to produce exhaustive libraries of all types of scientific time-series data, we have attempted to be as comprehensive and even-handed as possible.
However, emphasis was inevitably given to time series in publicly-available repositories, and outputs from time-series models that were relatively simple to implement.
A list of time series used in this work, including descriptions and references, is in the supplementary document {\it Time Series List}.

Methods for time-series analysis take on a variety of forms, from simple summary statistics to statistical model fits.
We implemented each such method as an algorithm: an {\it operation} that summarizes an input time series with a single real number.
Our library of over 9\,000 such operations quantify a wide range of time-series properties, including basic statistics of the distribution (e.g., location, spread, Gaussianity, outlier properties), linear correlations (e.g., autocorrelations, features of the power spectrum), stationarity (e.g., StatAv, sliding window measures, unit root tests, prediction errors), information theoretic and entropy measures (e.g., auto-mutual information, Approximate Entropy, Lempel-Ziv complexity), methods from the physical nonlinear time-series analysis literature (e.g., correlation dimension, Lyapunov exponent estimates, surrogate data analysis), linear and nonlinear model fits [e.g., goodness of fit and parameter values from autoregressive moving average (ARMA), generalized autoregressive conditional heteroskedasticity (GARCH), Gaussian Process, and state space models], and others (e.g., wavelet methods, properties of networks derived from time series, etc.).
A large component of this work involved implementing existing time-series analysis methods (including publicly available packages and toolboxes) in the form of operations.
In some cases it was necessary to formulate new types of operations that appropriately summarized the outputs of existing methods, and in other cases we developed new, qualitatively different types of operations, that are introduced for the first time in this work (see Sec.~S1.1.2 of the SI).
The collection is inevitably incomplete, and indeed this operational framework is more suited to those methods that can be automated than more subtle types of analysis that require delicate hands-on experimentation by a time-series analysis expert.
However, in the process of collecting and implementing these methods, we made a concerted effort to incorporate and appropriately automate as many distinct types of scientific time-series analysis methods as possible.
The result is sufficiently comprehensive to achieve the range of successful results reported in this work.
Note that although we list over 9\,000 operations, this number includes cases for which a single method is repeated for multiple parameter values (e.g., calculating the autocorrelation function at 40 different time lags constitutes 40 different operations despite simply varying a single parameter of a single method); the number of conceptually distinct methods is significantly lower (one estimate arrives at approximately 1\,000 unique operations, cf. Sec.~S1.1.2).
A full list of operations developed for this work, including references and descriptions, is in the supplementary document {\it Operation List}.

\begin{figure*}
   \centering
	\includegraphics[width = 14.5cm]{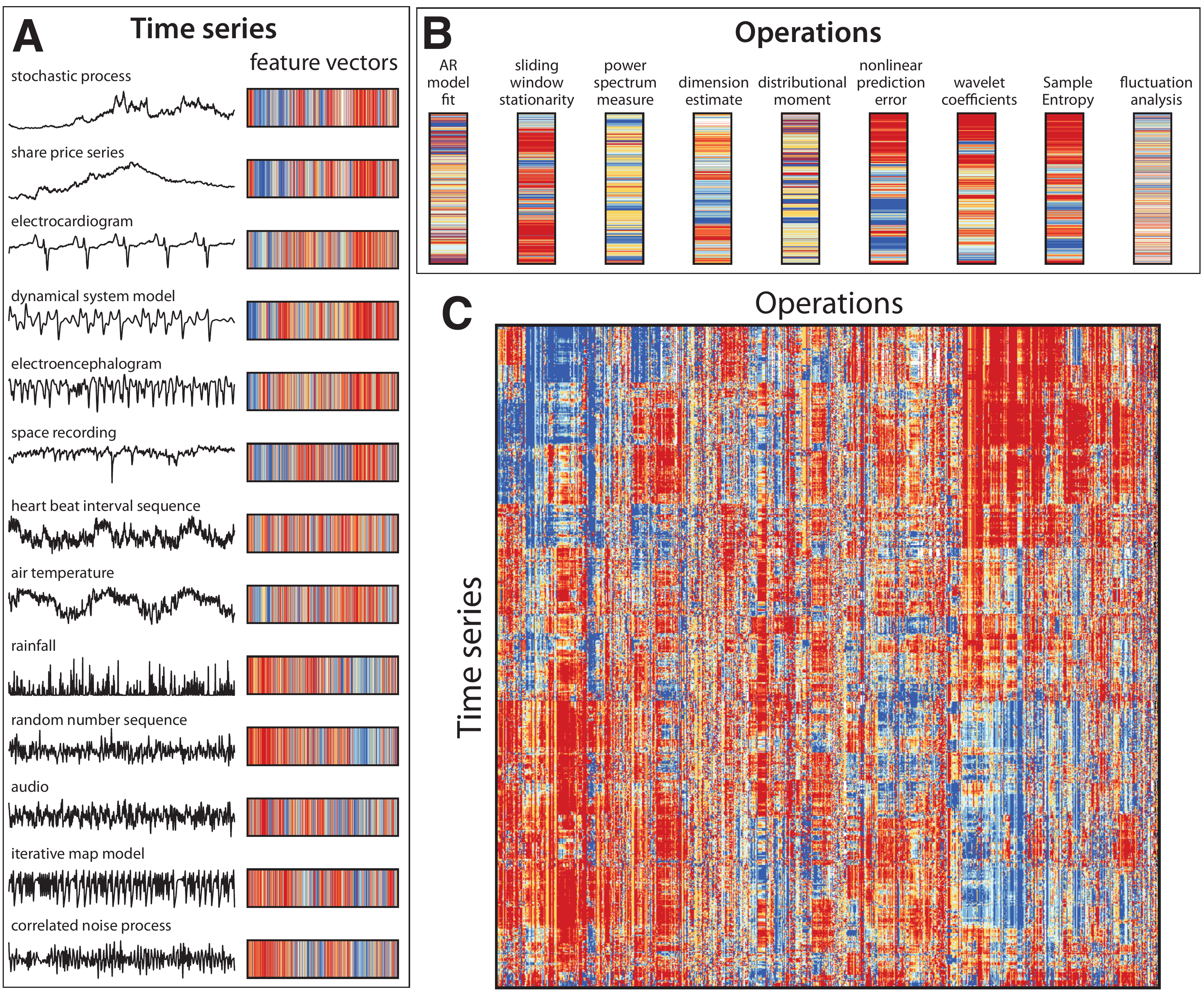}
   \caption{
	\textbf{A unified representation of time series and their analysis methods.}
	A fundamental component of this work, which involves applying large numbers of time-series analysis methods, or operations, to large time-series datasets, can be visualized as a data matrix, shown in \textbf{C}, where each column represents an operation and each row represents a time series.
	Elements of the matrix contain the (normalized) results of applying each operation to each time series, and are visualized using colour: from blue (low operation outputs) to red (high operation outputs).
	We note the similarity of this time-series data matrix to the DNA microarray used in biology, that simultaneously measures the expression levels under multiple conditions (our operations) of large numbers of genes (our time series).
	Rows and columns have been reordered using linkage clustering to place similar time series and similar operations close to one another (cf. Sec.~S1.3.2).
	Thus, time series (rows of the data matrix) are represented as feature vectors containing a large number of informative properties, illustrated in \textbf{A}, and operations (columns of the data matrix) are represented as feature vectors containing their outputs across a set of time series, illustrated in \textbf{B}.
	In this figure, we use an interdisciplinary set of 875 time series and 8\,651 well-behaved operations (cf. Sec.~1.1.5); the matrix in \textbf{C} has been resized to approximately square proportions for the purpose of visualization.
	The rich structure in this data matrix encodes relationships between different scientific time series and diverse types of analysis methods, relationships that are investigated in detail in this work.
	\label{fig:data_matrix}
}
\end{figure*}

A fundamental component of this work involves analyzing the result of applying a large set of operations to a large set of time series.
This computation can be visualized as a data matrix with time series as rows and operations as columns, as shown in Fig.~\ref{fig:data_matrix}C.
Each element of the matrix, $D_{ij}$, contains the output of an operation, $F_j$, applied to a time series, $\mathbf{x}_i$, so that $D_{ij} = F_j(\mathbf{x}_i)$.
Correspondingly, time series are represented as feature vectors containing measurements of an extensive range of their properties (Fig.~\ref{fig:data_matrix}A), and operations are represented as feature vectors containing their outputs across a time-series dataset (Fig.~\ref{fig:data_matrix}B).
In order to allow operations with different ranges and distributions of outputs to be compared meaningfully, we applied an outlier-robust sigmoidal normalizing transformation to the outputs of each operation, as described in Sec.~S1.2.2.
Time series are compared using Euclidean distances calculated between their feature vectors, and operations are compared using correlation-based distances measured between their outputs (either linear correlation-based distances to capture linear relationships or normalized mutual information-based distances to capture potentially nonlinear relationships, cf. Sec.~S1.3.1).
Thus time series are judged as similar that have many similar properties, and operations are judged as similar that have highly correlated outputs across a time-series dataset.
Note that when operations do not output a real number or are inappropriate (e.g., it is not appropriate to fit a positive-only distribution to non-positive data), these outputs are referred to as `special values' and are treated as missing elements of the data matrix that can be filtered out (cf. Sec.~S1.1.5).

\begin{figure*}
	\centering
		\includegraphics[width = 16cm]{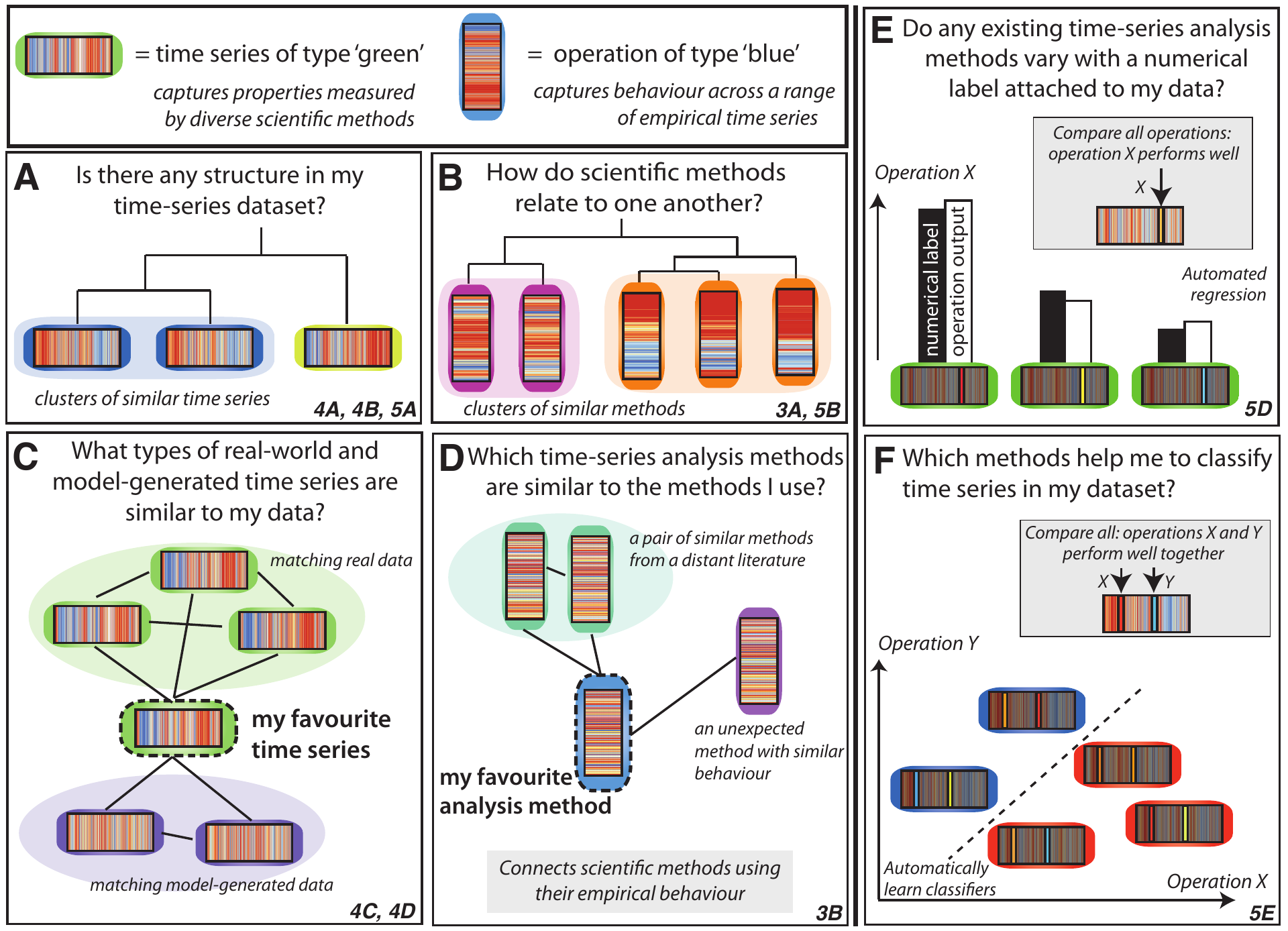}
	\caption{
	\textbf{Key scientific questions that can be addressed by representing time series by their properties (measured by many types of analysis methods) and operations by their behaviour (across many types of time-series data).}
	We show that this representation facilitates a range of versatile techniques for addressing scientific time-series analysis problems, that are illustrated schematically in this figure.
	The representations of time series (rows of the data matrix, Fig.~\ref{fig:data_matrix}A) and operations (columns of the data matrix, Fig.~\ref{fig:data_matrix}B), serve as empirical fingerprints, and are shown in the top panel.
	Coloured borders are used to label different classes of time series and operations, and other figures in this paper that explicitly demonstrate each technique are given in the bottom righthand corner of each subfigure.
	\textbf{A} Time-series datasets can be organized automatically, revealing the structure in a given dataset (cf. Figs.~4A, 4B, 5A).
	\textbf{B} Collections of scientific methods can be organized automatically, highlighting relationships between methods developed in different fields (cf. Figs.~3A, 5B).
	\textbf{C} Real-world and model-generated data with similar properties to a specific time series target can be identified (cf. Figs.~4C, 4D).
	\textbf{D} Given a specific operation, alternatives from across science can be retrieved (cf. Fig.~3B).
	\textbf{E} Regression: the behaviour of operations in our library can be compared to find operations that vary with a target characteristic assigned to time series in a dataset (cf. Fig.~5D).
	\textbf{F} Classification: operations can be selected based on their classification performance to build useful classifiers and gain insights into the differences between classes of labeled time-series datasets (cf. Fig. 5E).
	}
	\label{fig:empiricalfingerprints}
\end{figure*}

The rich structure in the data matrix shown in Fig.~\ref{fig:data_matrix}C, which combines the results of applying a wide range of scientific time-series analysis methods to a diverse set of time series, thus encapsulates interesting relationships between different ways of measuring structure in time series (columns), and relationships between data generated by different types of systems (rows).
For example, redundancy across analysis methods is indicated by a set of adjacent columns in the data matrix that display similar patterns (these operations exhibit similar behaviour across a great variety of different time series).
In this work, we introduce a range of simple techniques for extracting this kind of interesting and scientifically meaningful structure from data matrices, as illustrated schematically in Fig.~\ref{fig:empiricalfingerprints}.
In particular, we show that representing time series in terms of their measured properties, and analysis methods in terms of their behaviour on empirical data, can form a useful basis for answering the types of questions depicted in Fig.~\ref{fig:empiricalfingerprints}, including the ability to structure collections of data (Fig.~\ref{fig:empiricalfingerprints}A) and methods (Fig.~\ref{fig:empiricalfingerprints}B), find matches to particular time series (Fig.~\ref{fig:empiricalfingerprints}C) or methods (Fig.~\ref{fig:empiricalfingerprints}D), and perform time-series regression (Fig.~\ref{fig:empiricalfingerprints}E) or classification (Fig.~\ref{fig:empiricalfingerprints}F) automatically.
Our approach is unusual in that it is completely automated and uses no domain knowledge about the time series, or any information about the theoretical assumptions underlying the analysis methods: we simply use the empirical behaviour of methods and time series as a platform for comparison.
Using numerous examples from across science, we demonstrate that our libraries of time-series data and their analysis methods are sufficiently comprehensive for this approach to yield novel and scientifically meaningful results for a range of applications.

\section{Empirical structure} \label{sec:empirical_structure}
In this section we investigate the empirical structure of our annotated libraries of time series and their methods using the types of analysis depicted schematically in Figs. \ref{fig:empiricalfingerprints}A--D.
First, in Sec.~\ref{sec:empirical_structure_methods}, we show that the operations (columns of the data matrix in Fig.~\ref{fig:data_matrix}) can be organized in a meaningful way using their behaviour on a set of 875 different time series.
Then, in Sec.~\ref{sec:empirical_structure_timeseries}, we perform a similar treatment for time series (rows of the data matrix in Fig.~\ref{fig:data_matrix}).
We find that by judging operations and time series using a form of empirical fingerprint of their behaviour facilitates a useful means of comparing them.

\subsection{Empirical structure of time-series analysis methods}\label{sec:empirical_structure_methods}
First we analyze the structure in our library of time-series analysis operations when applied to a representative interdisciplinary set of 875 real-world and model-generated time series (a selection of time series in this set are illustrated in Fig.~\ref{fig:data_matrix}A).
This carefully controlled set of 875 time series was selected to encompass the different types of scientific time series as even-handedly as possible (for example, the full time-series library contains a large number of ECGs and rainfall time series; by using a controlled set we avoid emphasizing the behaviour of operations on these particular classes of time series that happen to be more numerous in our library, cf. Sec.~S2.1).
Operations that returned less than 20\% special-valued outputs on this set of time series were analyzed: a collection of 8\,651 operations.
In this section we show that the behaviour of operations on these 875 diverse scientific time series becomes a useful form of empirical fingerprint for them that groups similar types of operations and distinguishes very different types of operations.

\begin{figure*}
	\centering
		\includegraphics[width = 16cm]{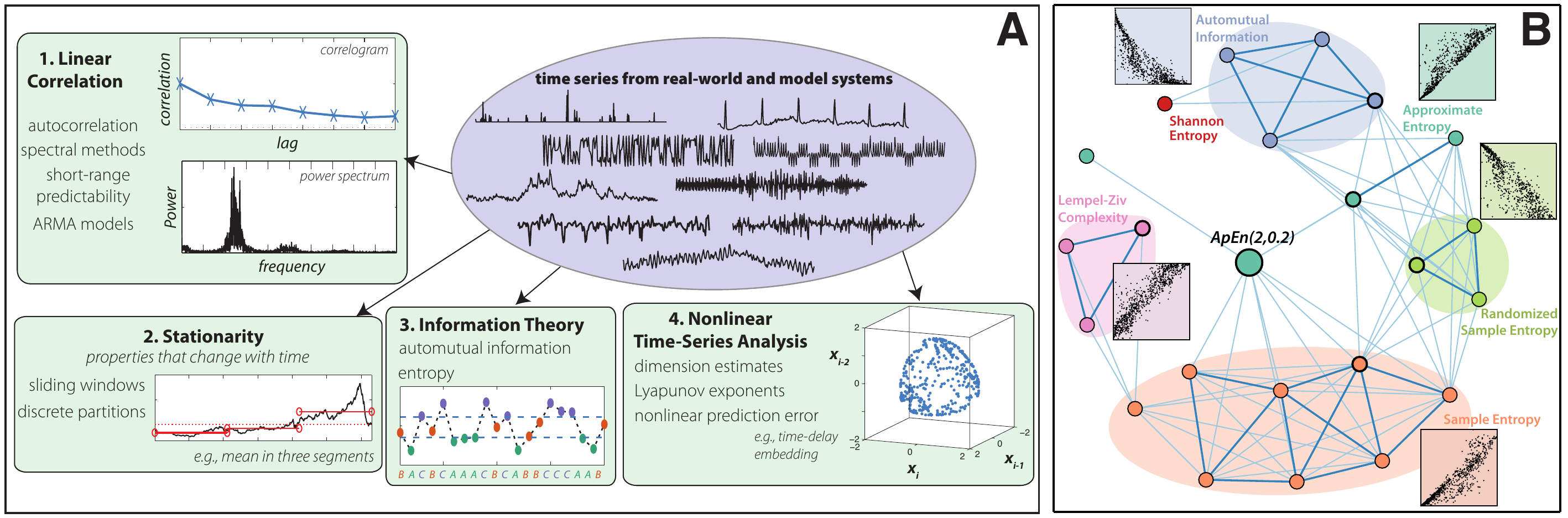}
	\caption{
	\textbf{Structure in a library of 8\,651 time-series analysis operations.}
	\textbf{A} A summary of the four main classes of operations in our library, as determined by a $k$-medoids clustering, reflects a crude but intuitive overview of the time-series analysis literature.
	\textbf{B} A network representation of the operations in our library that are most similar to the Approximate Entropy algorithm, ApEn(2,0.2) \cite{Pincus91}, which were retrieved from our library automatically.
	Each node in the network represents an operation and links encode distances between them (computed using a normalized mutual information-based distance metric, cf. Sec.~S1.3.1).
	Annotated scatter plots show the outputs of ApEn(2,0.2) (horizontal axis) against a representative member of each shaded community (indicated by a heavily outlined node, vertical axis).
	Similar pictures can be produced by targeting any given operation in our library, thereby connecting different time-series analysis methods that nevertheless display similar behaviour across empirical time series.
	\label{fig:op_clustering}
	}
\end{figure*}

We used clustering \cite{Hastie09} to uncover structure in these 8\,651 operations automatically (as in Fig.~\ref{fig:empiricalfingerprints}B).
Clustering can be performed at different resolutions to produce different overviews of the time-series analysis literature.
For example, clustering the operations into four groups using $k$-medoids clustering \cite{Hastie09} is depicted in Fig.~\ref{fig:op_clustering}A, and reflects a crude but intuitive summary of the types of methods that scientists have developed to study time series.
Although highly simplified, the result shows that we can begin to organize an interdisciplinary methodological literature in an automatic, data-driven way.

Next we ask a more subtle question: how many different types of operations are required to provide a good summary of the rich diversity of behaviours exhibited by the full set of 8\,651 scientific time-series analysis operations?
Clustering at finer levels allows us to probe this, and to construct reduced sets of operations that efficiently approximate the behaviours in the full library by eliminating methodological redundancy.
The quality of such reduced sets of operations is quantified using the {\it residual variance} measure, $1 - R^2$, where $R$ is the linear correlation coefficient measured between the set of distances between time series in a reduced space, and those in the full space \cite{Tenenbaum00}.
Using $k$-medoids clustering for a range of cluster numbers, $k$, we found that a reduced set of 200 operations provides a very good approximation to the full set of 8\,651 operations, with a residual variance of just 0.05 (cf. Sec.~S2.2). 
These 200 operations thus form a concise summary of the different behaviours of time-series analysis methods applied to scientific time series, and draw on techniques developed in a range of disciplines, including autocorrelation, automutual information, stationarity, entropy, long-range scaling, correlation dimension, wavelet transforms, linear and nonlinear model fits, and measures from the power spectrum (cf. supplementary document {\it 200 Operations}).
Constraints on the dynamics observed in empirical time series thus allows us to exploit redundancy in scientific methods acting on them.
In Sec. \ref{sec:empirical_structure_timeseries}, this reduced set of operations is used to organize our time-series library.

As well as analyzing the overall structure in our library of operations, it is also informative to explore the local structure surrounding a given target operation (as depicted in Fig. \ref{fig:empiricalfingerprints}D).
In this way, the behaviour of any given operation can be contextualized with respect to alternatives from across the time-series analysis literature, including methods developed in unfamiliar fields, or in the distant past.
An example is shown in Fig.~\ref{fig:op_clustering}B for the Approximate Entropy algorithm, ApEn(2,0.2), a `regularity' measure that has been applied widely \cite{Pincus91}.
In Fig.~\ref{fig:op_clustering}B, nodes in the network are the most similar operations to ApEn(2,0.2) in our library, and links indicate their similarities (using a mutual information-based distance metric to capture correlations across the set of 875 time series, cf. Sec.~S1.3.1). 
The network contains different communities of similar operations, including methods based on Sample Entropy, Lempel-Ziv complexity, automutual information, Shannon entropy, and other Approximate Entropies.
Inset scatter plots show the relationships between ApEn(2,0.2) and the other operations across different types of scientific time series.
We thus discover that, even when mixed with 8\,650 different operations, related families of entropy measures are retrieved automatically and organized meaningfully by comparing their behaviour across a controlled set of 875 time series.
Similar pictures can be produced straightforwardly for any operation in our library, as is done for fluctuation analysis and singular spectrum analysis, for example, in the SI (Sec.~S2.4). 

We thus find that representing time-series analysis methods using an empirical fingerprint of their behaviour across 875 different time series provides a useful means of comparing and structuring a methodological literature.
Across the scientific disciplines, there exists a vast number of time-series analysis methods, but no framework with which to judge whether progress is really being made through the continual development of new types of methods.
By comparing their empirical behaviour, the techniques demonstrated above can be used to connect new methods to alternatives developed in other fields in a way that encourages interdisciplinary collaboration on the development of novel methods for time-series analysis that do not simply reproduce the behaviour of existing methods.

\subsection{Empirical structure of time series}\label{sec:empirical_structure_timeseries}
\begin{figure*}
	\centering
		\includegraphics[width = 12cm]{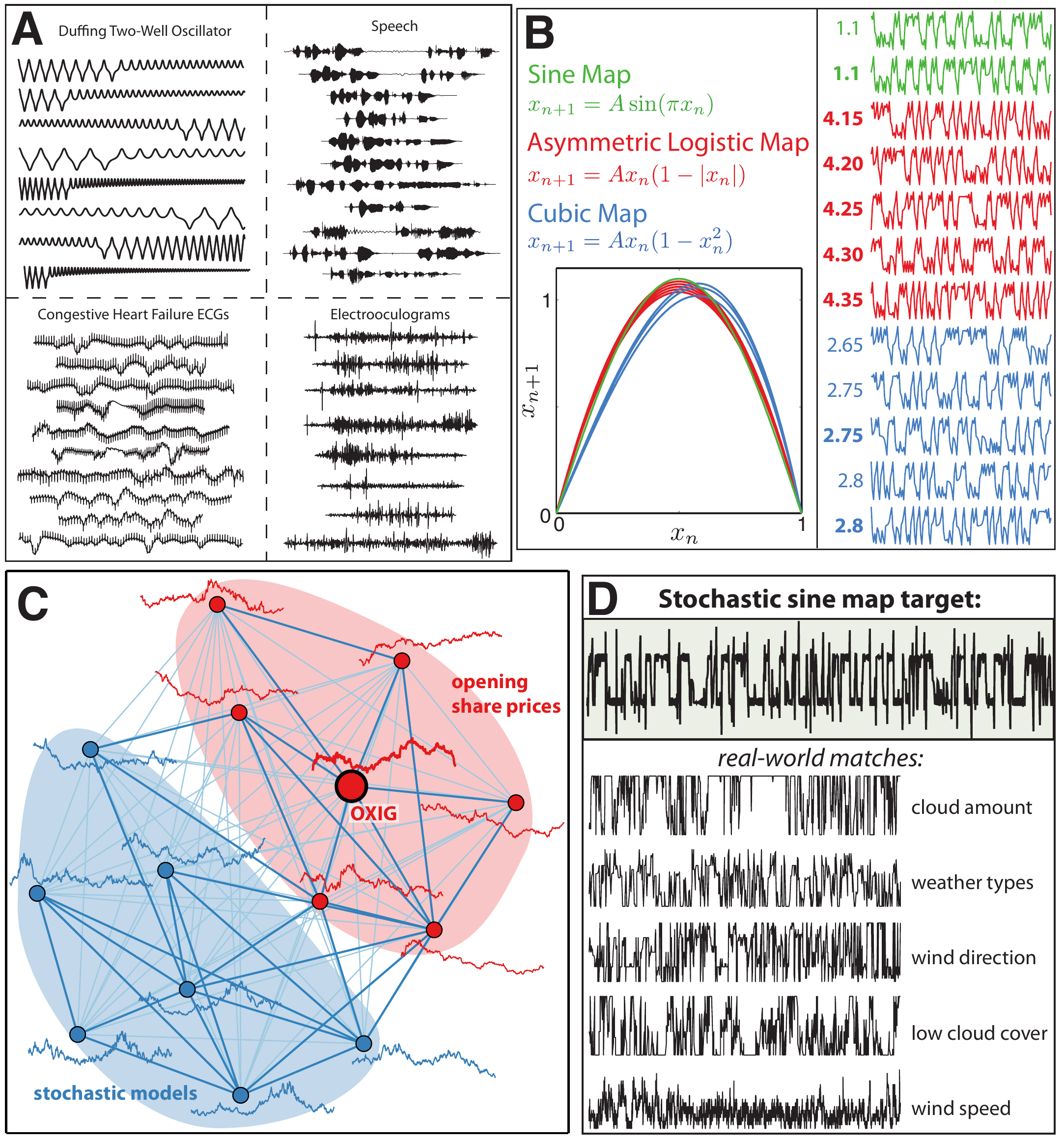}
	\caption{
	\textbf{Structure in a large, diverse library of scientific time series.}
	Representing time series using an empirical fingerprint containing 200 of their measured properties, a diverse collection of 24\,577 time series was clustered into 2\,000 groups.
	\textbf{A} Most clusters formed in this way are homogenous groups of time series of a given real-world or model system; four such examples are shown.
	\textbf{B}
	A time-series cluster is plotted that contains time series generated by three different iterative maps with parameters that specify a common recurrence relationship.
	Time-series segments of 150 samples are plotted and labeled with the parameter $A$ of the map that generated them (bold labels indicate 5\,000-sample time series, the others are 1\,000 samples long).
	Recurrence relationships, $x_{n+1}(x_n)$, are plotted and are similar for all time series.
	\textbf{C}
	An opening share price series for {\it Oxford Instruments} (labeled OXIG) is targeted; the most similar real-world time series are opening share prices of other stocks (red nodes), and the most similar model-generated time series are from stochastic differential equations (blue nodes).
	Links in the network represent similarities between the time series judged using Euclidean distances, $d$, between their normalized feature vectors (darker, thicker links have $d < 3$, cf. Sec.~S1.3.1). 
	\textbf{D} The most similar real-world time series to a Stochastic Sine Map target are meteorological processes that display qualitatively similar `noisy switching' dynamics.
	\label{fig:ts_clustering}
	}
\end{figure*}

Above we studied the structure in our library of scientific operations using their behaviour on empirical time series; in this section we analyze a collection of 24\,577 time series in a similar way.
Time series are compared using the set of 200 representative operations formed above to measure their properties, providing an empirical means of comparing them and addressing useful scientific questions (as depicted in Figs.~\ref{fig:empiricalfingerprints}A, C).
We restricted our clustering analysis to this set of 24\,577 time series, which is an appropriate subset of the full library of 38\,190 time series that removes very short time series (of less than 1\,000 samples) and filters over-represented classes of time series (as explained in Sec.~S3.2). 
Finding structure in a set of time series of different lengths and measured in different ways from different systems is a major challenge \cite{Wang12}; we show that this empirical fingerprint of 200 diverse time-series analysis operations facilitates a meaningful comparison of scientific time series.
First, we formed 2\,000 clusters of time series from our library (using complete linkage clustering, cf. Sec.~S3.2). 
Despite the size of our library and the diversity of time series contained in it, most clusters grouped time series measured from the same system; some examples are shown in Fig.~\ref{fig:ts_clustering}A (more examples are in Sec.~S3.2.1). 
Since the clustering is done by comparing a wide range of time-series properties, clusters appear to group time series according to their dynamics, even when they have different lengths.
Some clusters contained time series generated by different systems, such as the cluster illustrated in Fig.~\ref{fig:ts_clustering}B, that contains time series generated from three different iterative maps: the {\it Cubic Map}, the {\it Sine Map}, and the {\it Asymmetric Logistic Map} \cite{Sprott03}.
Although this cluster contains time series of different lengths and generated from different maps, it groups outputs from each map with parameters that specify a similar recurrence relationship, as shown in the inset plot of Fig.~\ref{fig:ts_clustering}B.
This cluster therefore distinguishes a distinct and meaningful class of dynamical behaviour from a large library of time series (as do many other clusters, cf. Sec.~S3.2.2). 
Conventional measures of time-series similarity are often based on distances measured between the time-series values themselves and are hence restricted to sets of time series of a fixed length \cite{Wang12}.
Representing time series by a diverse range of their properties is clearly a powerful alternative that captures important dynamical behaviour in general collections of scientific time series.

Our reduced representation of time series also allows us to retrieve a local neighborhood of time series with similar properties to any given target time series (note that this search is done across the full library of 38\,190 time series).
This automated procedure can be used to relate real-world time series to similar, model-generated time series in a way that suggests suitable families of models for understanding real-world systems (as depicted in Fig.~\ref{fig:empiricalfingerprints}C).
An example is shown in Fig.~\ref{fig:ts_clustering}C, where real-world and model-generated matches to a series of opening share prices for {\it Oxford Instruments} (OXIG) are plotted as a network.
Real-world matches are other opening share price series, and model-generated matches are outputs from stochastic differential equations (SDEs).
Indeed, the form of SDE models suggested by this set of matches, e.g., $dX_t = \mu X_t dt + \sigma X_t dW_t$, is the same as geometric Brownian motion, which is used in financial modeling \cite{Samuelson65} (where $X_t$ represents the time series, $W_t$ denotes a Wiener process, and other variables are parameters).
Matches to SDE models with slightly different forms, e.g., $dX_t = a(b-X_t)dt + \sigma\sqrt{X_t}dW_t$, suggest that other models with particular parameter values can also reproduce many properties of the target share price series that are not unique to the geometric Brownian motion model.
Although this is an extremely crude alternative to conventional time-series modeling, it nevertheless allows real-world time series to be linked to relevant model systems in a completely automated and data-driven way.
Analogous insights were gained for rainfall patterns, astrophysical recordings, human speech recordings, and others in Sec.~S3.3. 

Applying the same method in `reverse', we also targeted time series generated by models and retrieved real-world time series with similar dynamics.
For example, in Fig.~\ref{fig:ts_clustering}D, we targeted a time series generated by the stochastic sine map model, which has a fixed probability of additive uniformly-distributed noise at each time step that can switch the system between two stable limit cycles \cite{Freitas09}.
As expected, the closest model-generated matches were other stochastic sine map time series generated using the same parameters as the target (not shown here, cf. Sec.~S3.3.2), while real-world matches were meteorological processes that exhibit the same qualitative `noisy switching' dynamics (Fig. \ref{fig:ts_clustering}D). 
Thus, the dynamics specified by the stochastic sine map model were explicitly linked to that of relevant real-world meteorological processes automatically, suggesting that this type of stochastic switching mechanism may capture some of the properties of these meteorological systems.
Using this simple, general method for connecting real-world and model dynamics, we obtained similar results using noise-corrupted sine waves and self-affine time series in Sec.~S3.3.2. 

We have introduced large, annotated collections of time series and their methods from across the scientific disciplines and used the behaviour of each applied to the other to analyze structure in them.
Although representing time series and their methods in terms of their empirical behaviour might seem like an unusual idea that might not yield a particularly meaningful results, we found that it indeed provides a powerful, data-driven means of investigating relationships between them.
In particular, representing time series using 200 operations and operations using 875 time series appears to be sufficient to organize them meaningfully.
As illustrated in Figs.~\ref{fig:empiricalfingerprints}A--D, the methods we introduce exploit this unified representation to connect individual pieces of time-series data to other time series with similar properties, and particular time-series analysis methods to alternatives with similar behaviour.
The examples shown here and in the comprehensive SI demonstrate the general applicability of these techniques to a broad range of time-series analysis problems.

\section{Applications} \label{sec:applications}
In this section, we show how our library of operations can be used to provide new insights into the analysis of specific time-series datasets.
We employ the types of analysis demonstrated above, including organizing sets of time series (depicted in Fig.~\ref{fig:empiricalfingerprints}A) and methods (Fig.~\ref{fig:empiricalfingerprints}B), and also introduce new techniques for selecting useful operations for classification and regression tasks automatically (Figs.~\ref{fig:empiricalfingerprints}E,~F).
The broad utility of our highly comparative approach to specific scientific applications is shown using a wide range of scientific case studies.
Despite comparing across thousands of operations, we find that the selection of useful operations can be done in a statistically controlled manner using a multiple hypothesis testing framework described in Sec.~S1.3.6. 
A detailed analysis of each case study, including information about the dataset, statistical tests, interpretations of selected operations, and further comparisons to the existing literature, can be found in Sec.~S4 (time-series regression) and Sec.~S5 (time-series classification) of the SI. 

\begin{figure*}
	\centering
		\includegraphics[width = 16.0cm]{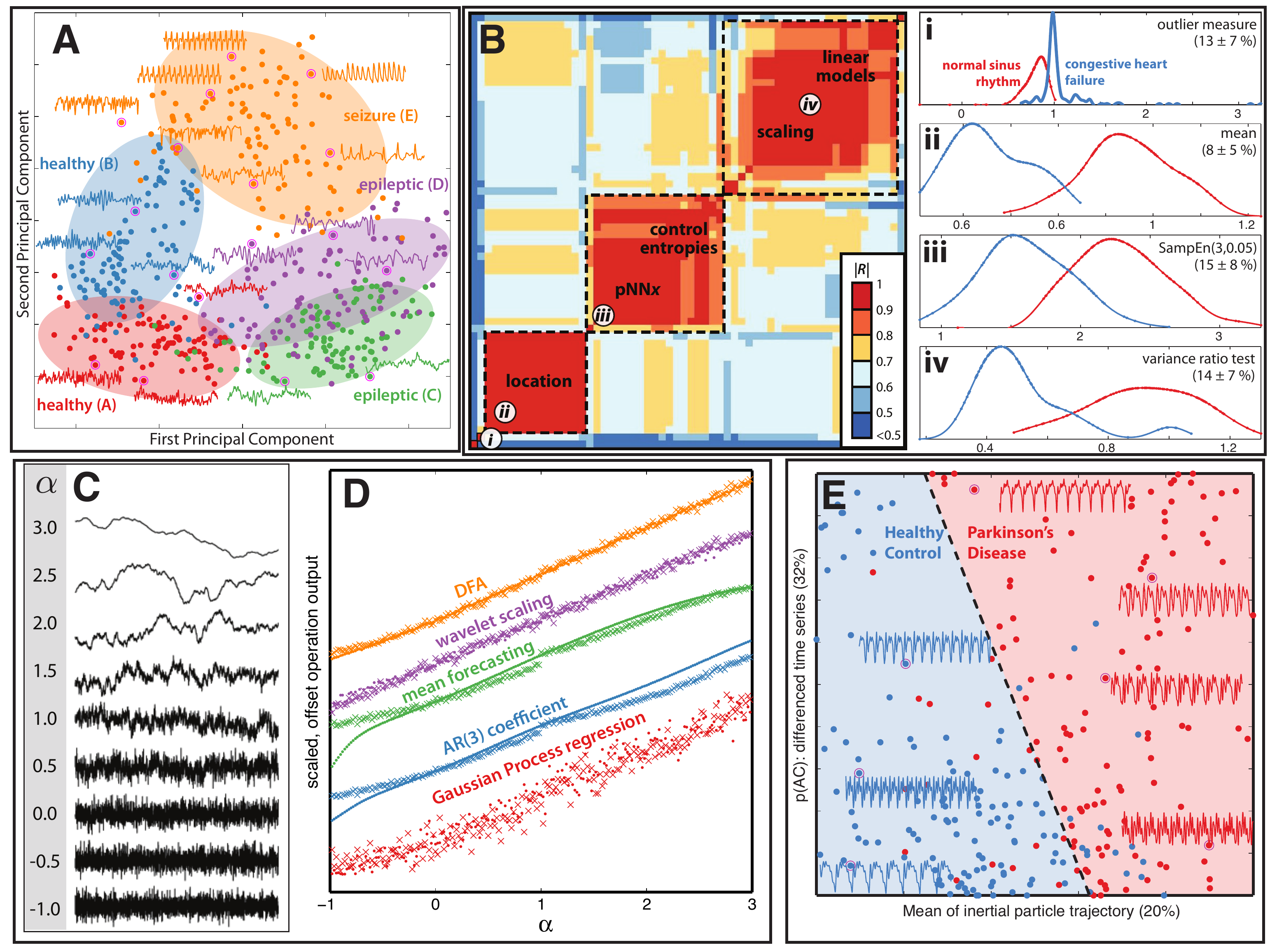}
	\caption{
	\textbf{Highly comparative techniques for time-series analysis tasks.}
	We draw on our full library of time-series analysis methods to: structure datasets in meaningful ways (\textbf{A}) and retrieve and organize useful operations for classification (\textbf{B} and \textbf{E}) and regression tasks (\textbf{C} and \textbf{D}).
	\textbf{A} Five classes of EEG signals are structured meaningfully in a two-dimensional Principal Components space of our library of operations.
	\textbf{B} Pairwise linear correlation coefficients measured between the sixty most successful operations for classifying congestive heart failure and normal sinus rhythm RR interval series.
	Clustering reveals that most operations are organized into one of three groups (indicated by dashed boxes).
	Distributions of selected operations, labeled \textbf{i}, \textbf{ii}, \textbf{iii} and \textbf{iv} in the correlation matrix, are plotted for the two classes, including the mean $\pm$ standard deviation of their 10-fold cross-validation misclassification rate using a linear (threshold) classifier.
	\textbf{C} Segments of self-affine time series generated with scaling exponents in the range $-1 \leq \alpha \leq 3$.
	\textbf{D} Five selected operations with outputs that vary approximately linearly with $\alpha$ for self-affine time series generated by the Fourier filtering method (dots), and the random midpoint displacement method (crosses).
	\textbf{E} A linear classifier that combines the outputs of two different operations distinguishes Parkinsonian speech recordings (red) from those of healthy controls (blue) with a mean 10-fold cross-validation misclassification rate of 14\%; the mean cross-validation misclassification rate of each individual operation is given in parentheses.
	Some sample segments of time traces are annotated to the plot.
	\label{fig:applications}
	}
\end{figure*}

\subsection{EEG recordings} \label{sec:eegs}
In the first case study, we show how the collective behaviour of our library of operations can be used to organize different types of electroencephalogram (EEG) time series (cf. Fig.~\ref{fig:empiricalfingerprints}A), and we assess whether progress is being made in a literature concerned with the classification of epileptic seizure EEGs (cf. Fig.~\ref{fig:empiricalfingerprints}F).
The dataset contains 100 EEG signals from each of five classes: set {\it A} (healthy, eyes open), set {\it B} (healthy, eyes closed), set {\it C} (epileptic, not seizure, recorded from opposite hemisphere of the brain to epileptogenic zone), set {\it D} (epileptic, not seizure, recorded within the epileptogenic zone), and set {\it E} (seizure) \cite{Andrzejak01}.
The two-dimensional Principal Components projection of the dataset in the space of well-behaved operations is shown in Fig.~\ref{fig:applications}A.
This representation uses no knowledge of the class labels attached to the data, but simply organizes the time series according to their measured properties using our large and diverse library of operations.
The dataset is structured in a way that is consistent with its known class structure: seizure time series (set {\it E}) are particularly distinguished, and the two healthy (sets {\it A} and {\it B}) and the two epileptic (sets {\it C} and {\it D}) sets are close in this space.
Drawing on a large literature of time-series analysis operations using lower-dimensional representations can evidently uncover useful structure in time-series datasets, and indeed the same approach was found to be informative for a range of other systems (including periodic and noisy signals in Sec.~S5.1.2, seismic signals in Sec.~S5.2.3, emotional speech recordings in Sec.~S5.3.3, and RR intervals in Sec.~S5.6.2). 

This dataset has previously been used to build classifiers that distinguish healthy EEGs from seizures using sets {\it A} and {\it E}.
For example, one study reports a support vector machine classifier using operations derived from the discrete wavelet transform with a classification accuracy of at least 98.75\% using re-sampled subsegments of this dataset \cite{Subasi10}.
However, as shown in Fig.~\ref{fig:applications}A, these two types of signals are separated automatically in the two-dimensional Principal Components space of operations, indicating that this task is relatively straightforward when exploiting a wide range of time-series analysis methods.
Indeed, 172 different operations in our library each {\it individually} discriminated between healthy EEGs and seizures with a 10-fold cross-validation linear classification rate exceeding 95\%, with eight single operations exceeding 98.75\% (Sec.~S5.4.3). 
These most successful operations are derived from diverse areas of the time-series analysis literature, and provide interpretable insights into this classification problem: e.g., revealing that EEGs recorded during seizures have lower entropy, lower correlation dimension estimates, lower long-range scaling exponents, and distributions with a greater spread than those from healthy patients.
In contrast to existing research that tends to focus on developing increasingly complicated classifiers for this problem (cf. Sec.~5.4.3 for details), the comparative analysis performed here selects simple and interpretable methods for the task automatically. 
Without performing such a comparison, it is difficult to assess whether methodological progress is being made in a given literature, or whether many simpler alternative methods may outperform the current state of the art.

\subsection{Heart rate variability}\label{sec:hrv}
We applied our highly comparative techniques to the problem of distinguishing `normal sinus rhythm' and `congestive heart failure' heart-beat (RR) interval series.
Many existing studies have analyzed RR interval data, each reporting the usefulness of a particular method, or a small set of methods \cite{Malik96}.
In contrast, we show how a range of useful methods for this task can be selected from our library and organized in a way that synthesizes a large and disparate literature on the subject, as well as identifying promising new types of analysis techniques.
Our dataset contains 105 recordings from each class, with lengths ranging from 800 to 19\,900 samples (cf. Sec.~S5.6.1). 
Although the analysis could have been performed on a dataset of time series of a fixed length, here we show how informative operations can be retrieved even in the case where the time-series recordings are of very different lengths.

The 60 most successful operations for this task (those with a mean 10-fold classification rate exceeding 85\%), were retrieved from our library and are represented as a clustered pairwise similarity matrix in Fig.~\ref{fig:applications}B, where colour represents the linear correlation coefficient measured between the normalized outputs of all pairs of operations.
Most operations cluster into three main groups of behaviour: measures of location (e.g., mean, median), entropy/complexity estimates and {\it PNNx} measures \cite{Malik96}, and linear correlation-based methods (e.g., autoregressive model fits and power spectrum scaling).
Two other methods, in the bottom-left corner of the matrix in Fig.~\ref{fig:applications}B, display relatively unique behaviour on the dataset: a discrete wavelet transform-based operation, and an outlier-adjusted autocorrelation measure.
Example distributions from each of these families of successful operations are plotted in panels of Fig.~\ref{fig:applications}B: \textbf{i} an outlier measure returns the ratio of lag-3 autocorrelations of the time series before and after removing 10\% of outliers, \textbf{ii} the mean, \textbf{iii} the Sample Entropy \cite{Richman00}, SampEn(3,0.05), and \textbf{iv} a variance ratio hypothesis test developed in the Economics literature \cite{Cecchetti94}.
Despite being selected in a completely automated way, each selected operation provides an understanding of the dataset by contributing an interpretable measure of structural difference between the two classes of RR intervals.
For example, the well-known results that normal sinus rhythm series tend to have longer inter-beat intervals (higher mean, Fig.~\ref{fig:applications}B\textbf{ii}) and greater entropy (higher Sample Entropy, Fig.~\ref{fig:applications}B\textbf{iii}) than congestive heart failure series.

This case study demonstrates the ability of our highly comparative approach to select and also organize useful methods for time-series classification tasks automatically, using their empirical behaviour.
The result provides interpretable insights into the differences between the known classes of time series.
This case study represents another example of how an interdisciplinary methodological literature can be structured in a purely data-driven way (cf. Fig.~\ref{fig:empiricalfingerprints}B).
However, unlike the treatment applied to general time-series analysis operations in Sec.~\ref{sec:empirical_structure_methods} above, here additional knowledge (in the form of class labels assigned to the data) is used to guide the selection of useful methods, which are then organized according to their behaviour on the data.
The resulting synthesis highlights similarities between different methods and distinguishes the novelty of others, and could be used to assess new contributions to the literature.
For example, a new scientific paper could introduce the variance ratio hypothesis test \cite{Cecchetti94}---a method originally developed in the Economics literature---as a novel measure for analyzing heart rate variability data.
However, as shown Fig.~\ref{fig:applications}B, the operation behaves like a range of simple linear model-based methods on this dataset, i.e., it is immediately clear that it is not actually measuring a new property of these time series but is simply reproducing the behaviour of these existing operations.
On the other hand, a new operation that we devised and introduce in this work, based on the impact of outliers on time-series autocorrelation properties, is distinguished as both unique and useful for this dataset (detailed information about all selected operations is in Sec.~S5.6.3). 
Referencing our comparative framework in this way can thus help to ensure that new methods actually represent new contributions to a given analysis literature.

\subsection{Self-affine time series}\label{sec:selfaffine}
Fluctuation analysis has attracted substantial attention in the statistical physics literature, and has been used to provide evidence for scale-invariance in a variety of real-world processes.
But do these conventional methods outperform alternative approaches, or do simpler, faster methods exist that display comparable (or even superior) performance?
We addressed this question by generating a synthetic dataset of self-affine time series and then searching our library (in a statistically controlled way, cf. Sec.~S1.3.6) for operations that vary linearly with their known scaling exponents (as depicted schematically in Fig.~\ref{fig:empiricalfingerprints}F). 
Self-affine time series can be characterized by a single scaling exponent, $\alpha$, according to $S(f) \propto f^{-\alpha}$, where $S$ is the power spectral density as a function of frequency, $f$ \cite{Peitgen88}.
Time series of 5\,000 samples were generated with scaling exponents uniformly distributed in the range $-1 \leq \alpha \leq 3$ by two different methods: (i) 199 time series generated using the {\it Fourier filtering method}, and (ii) 199 time series generated using the {\it random midpoint displacement method} \cite{Peitgen88} (cf. Sec.~S4.3), as shown in Fig.~\ref{fig:applications}C. 
In order to accurately characterize the scaling exponent, $\alpha$, of these time series, operations must combine local and global information to capture the self-similarity of the time series across multiple time scales.

A selection of five operations with the strongest linear correlations to $\alpha$ are plotted in Fig.~\ref{fig:applications}D.
As expected, many fluctuation analysis-based operations perform well on this task (cf. Sec.~S4.3), including the scaling exponent estimated using detrended fluctuation analysis (DFA) \cite{Peng95a}, and a wavelet-based alternative (labeled `DFA', and `wavelet scaling' in Fig.~\ref{fig:applications}D, respectively). 
However, other types of operations that are not based on fluctuation analysis also exhibit strong linear correlations with $\alpha$: the autocorrelation of residuals from a local mean forecaster (labeled `mean forecaster' in Fig.~\ref{fig:applications}D), the first-order coefficient of an autoregressive AR(3) model fitted to the time series (labeled `AR(3) coefficient'), and the mean log hyperparameter of the squared exponential length scale from a Gaussian Process regression on local segments of the time series (labeled `Gaussian Process regression').

Thus, as well as confirming the utility of methods based on fluctuation analysis to estimate the scaling exponent of self-affine time series, we also discovered a surprising selection of other useful methods that capture the known variation in $\alpha$ by combining local and global structure in various ways.
Many of these alternative methods require significantly less computational effort and can be updated iteratively, making them more suited to real-time applications than the slightly more accurate but also more computationally intensive fluctuation-based methods.
Our highly comparative approach to time-series regression, that selects relevant scientific operations based on their empirical performance, can thus be used to find fast approximations to traditional methods for real applications involving finite, noisy time series.
This general approach could also be used to select methods that help predict important diagnostic quantities assigned to physiological recordings, such as predicting the stage of sleep from an EEG recording or the arterial pH of a baby from its fetal heart rate time series recorded during labor.
Additional examples are presented in the SI, where we successfully selected interpretable estimators of the variance of white noise added to periodic signals (Sec.~S4.1) and the Lyapunov exponent of Logistic Map time series (Sec.~S4.2). 
The success of our highly comparative approach relies on our library of operations being sufficiently comprehensive to produce useful and informative results.

\subsection{Parkinsonian speech}\label{sec:parkinsons}
This final case study involves the particularly challenging task of distinguishing stationary phonemes recorded from patients with Parkinson's disease from those of healthy controls \cite{Little09}.
Rather than implementing a specific set of standard speech analysis techniques, the structure in the data was used to select appropriate operations from our general library.
Using the procedure illustrated in Fig. \ref{fig:empiricalfingerprints}F, we found that our operation library is rich enough to construct useful classifiers for this difficult task.
The dataset contains 127 speech recordings from Parkinsonian patients and 127 speech recordings from healthy controls, with recording lengths ranging from 50\,ms to 1.25\,s (cf. Sec.~S5.5). 

We found that many of the most successful operations in our library for classifying Parkinsonian speech were closely related to existing measures from speech analysis (e.g., the `jitter' summary statistic \cite{Little09}), and we also identified some new techniques (cf. Sec.~S5.5.2). 
However, rather than restricting ourselves to individual operations, we also used forward feature selection to construct classifiers that combine pairs of complementary operations to achieve improved out-of-sample classification performance (using appropriate partitions of data into training and test sets, as described in Sec.~S1.3.5). 
An example two-feature linear discriminant classifier is shown in Fig.~\ref{fig:applications}E, and combines a new operation introduced by us in this work, that simulates an inertial particle that experiences an attractive force to the time series, with another operation that constructs a symbolic string from incremental differences of the time series using a three-letter alphabet (`A', `B', `C'), and returns the probability of the word `AC'.
The two operations complement one another and the resulting classifier is simple, interpretable, and has a mean 10-fold cross-validation misclassification rate of 14\%, which is comparable to the state-of-the-art in this literature (that typically treats unbalanced datasets of fixed-length time series \cite{Little09}, cf. Sec.~S5.5.2). 
Using our method, multi-feature classifiers for time series classification are constructed automatically, require no domain knowledge of the mechanisms underlying the time series, and can be extended to classifiers containing three or more operations straightforwardly (although for this dataset we found minimal out-of-sample improvement in classification rate on adding more than two features, cf. Sec.~S5.5.3). 

To further demonstrate the applicability of our methods to diverse types of time-series recordings, we analyzed two additional datasets in the SI, where an unknown seismic recording was classified as an explosion rather than earthquake, consistent with previous studies (Sec.~S5.2), and competitive multi-feature classifiers were constructed for distinguishing seven different classes of emotional content in human speech recordings (Sec.~S5.3). 
Thus, despite using extremely simple statistical learning techniques (including linear classifiers, forward feature selection, linkage and $k$-medoids clustering, and Principal Components Analysis), we are able to contribute meaningful results to a wide variety of scientific time-series analysis problems.
These simple methods have the advantage of being transparent and producing readily interpretable results to demonstrate our approach; more sophisticated methods should yield improved results and can be explored in future work.
On several occasions we also found that new operations developed by us in this work were amongst the most useful operations for various analysis tasks, including the outlier autocorrelation measure selected for classifying heat beat interval data in Sec. \ref{sec:hrv} and the particle trajectory-based operation for Parkinsonian speech in Sec. \ref{sec:parkinsons}.
This highlights the benefit of being creative in the development of new operations for our library, as useful  operations are selected based on their performance on real datasets.
The results of this section demonstrate that our library of operations is sufficiently comprehensive to be broadly useful in guiding the selection of methods for scientific time-series analysis tasks.

\section{Conclusions}\label{sec:conclusions}
In summary, we have shown that interesting and scientifically meaningful structure can be discovered automatically in extensive annotated collections of time-series data and time-series analysis methods from a wide selection of theoretical and empirical literatures.
Relationships are determined by comparing empirical behaviour: the outputs of the methods applied to data, and the properties of the data as measured by the methods.
Representing time series and operations in this way turns out to provide a powerful framework for organizing general collections of time series and operations.
By unifying a previously disjoint interdisciplinary literature, we thus motivate a complementary and highly comparative approach to time-series analysis that provides insights into the properties of time series studied in science and the techniques that scientists have developed to study them.

Using our framework, data analysts can now readily ask new types of questions of their time-series data and methods, as depicted in Fig.~\ref{fig:empiricalfingerprints}.
In Sec.~\ref{sec:empirical_structure_methods}, we demonstrated how time-series analysis operations can be represented using their outputs across a controlled set of 875 real and model-generated time series from across the sciences.
This form of empirical fingerprinting allowed us to structure a large and diverse library of scientific methods in a meaningful way.
The result provides time-series analysts a new means of understanding the methods they use to analyze their data as part of a wider scientific context.
For example, data analysts can now investigate relationships between the set of methods familiar to them and extensive libraries of alternative methods that may have been developed in different disciplinary contexts.
Furthermore, new methods for time-series analysis can now be compared to the existing literature (as represented in our library) to check for redundancy and hence help ensure that new methods for time-series analysis really constitute advances and do not simply reproduce existing behaviour.
In Sec.~\ref{sec:empirical_structure_timeseries}, using the output of 200 diverse operations as a form of empirical fingerprint for a time series allowed us to identify meaningful clusters from a large and diverse library, and to retrieve meaningful matches to a given target time series.
For example, on receiving a new dataset a time-series analyst may wish to understand what key structures exist in it, or what other types of real-world and model-generated time series have similar types of dynamics: this can now be achieved in an automated, data-driven fashion.
By structuring datasets and connecting them to relevant types of real-world and model systems, the results provide an interdisciplinary context for the problem that can be used to guide a more focused analysis.

As well as providing useful tools for understanding the structure in large collections of time series and their methods, in Sec.~\ref{sec:applications} we showed how additional knowledge about particular datasets can be incorporated to automate the selection of useful analysis methods.
Because each operation provides an interpretable measure of some kind of structure in the time series, selecting operations in this way yields insights into the most informative properties of the data and how they vary across the labeled classes.
The operations are selected according to their behaviour, and the result connects a range of ways of thinking about structure in time series to a given analysis problem.
Using EEG seizure data (Sec.~\ref{sec:eegs}) we showed how difficult it can be to assess existing studies that quote classification rates without comparison to alternative methods and demonstrated how our library of operations can be used to perform this comparison.
In Sec.~\ref{sec:hrv}, we retrieved and organized operations that help to distinguish healthy and congestive heart beat intervals by comparing the behaviour of all operations on the data.
Each operation provides an interpretable measure of structural difference between the two classes, and the organization reveals which groups of operations exhibit similar behaviour on the dataset and which are unique (Fig.~\ref{fig:applications}B).
These results suggest that only a reduced number of operations need to be measured to characterize and classify the heart beat interval data, information that could be used by a domain expert to guide the selection of a reduced set of unique analysis operations.
Further investigation of the selected operations by a domain expert could also provide new insights into the dynamical structure of heart beat intervals for healthy patients and for various pathologies.
A similar approach was used to select operations that most accurately predict the scaling exponent, $\alpha$, of self-affine time series in Sec.~\ref{sec:selfaffine}, and we also demonstrated how forward feature selection can be used to construct interpretable multi-feature classifiers for identifying Parkinsonian speech recordings.
The success of our approach in each case relied on our library of operations being sufficiently comprehensive.
Thus, although is far from complete, our library appears to contain enough diversity to provide good classification performance and contribute useful insights into a range of time-series analysis problems.
The results could be improved further in the future by refining and growing the library with methods contributed by time-series analysis experts.

We now revisit the analogy to the high throughput analysis of the DNA microarray in biology, which is used routinely to guide more focused research efforts in that field.
While time-series analysis has traditionally focused on the use of specific methods and models motivated by domain knowledge, the highly comparative techniques introduced in this work represent a similarly powerful complement to this approach.
As vast quantities of time-series data continue to be recorded, and new methods for their analysis are developed, this highly comparative approach will allow us to make sense of this large and complex resource and contribute to directing progress in an inherently interdisciplinary field.
The results are wide-reaching, from the diagnosis of pathologies in medical recordings to the detection of anomalies in an industrial process or on an assembly line.
We hope that our work, while stimulating theory, adds to the experimental and empirical aspect of the study of time-series data.
The Matlab source code for all operations used in this work can be obtained from the authors via the website \url{http://www.comp-engine.org/timeseries/}, and instructions of how our framework can be applied to new datasets is summarized in Sec.~S1.4 of the SI. 

\paragraph{Acknowledgements}
The authors would like to thank Summet Agarwal, Siddarth Arora, Andrew Phillips, Stephen Blundell, and Anna Lewis for valuable feedback and discussion on the manuscript, and David Smith for assistance with network visualization.
Nick Jones thanks the BBSRC and EPSRC BBD0201901, EP/H046917/1, EP/I005765/1 and EP/I005986/1.


\end{document}